\documentclass[11pt]{article}
\usepackage{amssymb}  
\usepackage{amsmath}
\usepackage{tikz}
\usetikzlibrary{arrows,shapes}
\usepackage{bm,bbm}
\usepackage{epsfig}
\usepackage{amsmath,mathtools}
\usepackage{cases}
\usepackage{fancybox}
\usepackage{graphicx}
\usepackage[ampersand]{easylist}
\usepackage{breqn}
\usepackage{amsmath}
\usepackage{amsthm}

\newtheorem*{parameter interpretation}{parameter interpretation}
\usepackage{float}
\usepackage{bigints}
\usepackage{relsize }
\usepackage{subfigure}
\usepackage{soul}
\usepackage{xcolor}
\usepackage{pdfpages}
\usepackage{cite}
\usepackage{array}
\usepackage{color,soul}
\usepackage{tikz}
\usepackage{listings}
\usepackage{booktabs}
\usepackage{authblk}
\usetikzlibrary{shapes.gates.logic.US,trees,positioning,arrows}

\usepackage[inner=3 cm,outer=3 cm,bottom=2 cm]{geometry}

\usepackage{setspace}
\title{Heritability Estimation in Matrix-Variate Mixed models- A Bayesian Approach}

\author{Najla Saad Elhezzani}
\affil{Department of Medical and Molecular Genetics, King's College London}
\affil{Department of Statistics, London School of Economics and Political Science}
\affil{Department of Statistics, King Saud University}
\begin{document}
\maketitle
{\fontsize{8}{8}\selectfont \tableofcontents}

\newpage 
\begin{abstract} 
\addcontentsline{toc}{section}{Abstract}
Since the emergence of genome-wide association studies (GWASs), estimation of the narrow sense heritability explained by common single-nucleotide polymorphisms (SNPs) via linear mixed model approaches became widely used. As in most GWASs, most of the heritability analyses are performed using univariate approaches i.e. considering each phenotype independently. 
\par In this study, we propose a Bayesian matrix-variate mixed model that takes into account the genetic correlation between phenotypes in addition to the genetic correlation between individuals which is usually modelled via a relatedness matrix. We showed that when the relatedness matrix is estimated using all the genome-wide SNPs, our model is equivalent to a matrix normal regression with matrix normal prior on the effect sizes. Using real data we demonstrate that there is a boost in the heritability explained when phenotypes are jointly modelled ($\sim 25-35\%$ increase). In fact based on their standard error, the joint modelling provides more accurate estimates of the heritability over the univariate modelling. Moreover, our Bayesian approach provides slightly higher estimates of heritability compared to the maximum likelihood method. On the other hand, although our method performs less well in phenotype prediction, we note that an initial imputation step relatively increases the prediction accuracy.
\end{abstract}

\section{Introduction}
Most genome-wide association studies (GWASs) as well as heritability estimations are conducted using univariate approaches, i.e. considering each phenotype independently, see for example [1]. This is because they are more computationally tractable and easier to interpret. For example, rejecting a null hypothesis of no association does not indicate which phenotypes are associated which requires a second stage analysis such as performing model comparison. However, many evaluations of significant GWASs found that SNPs appear to be associated with multiple, sometimes seemingly distinct phenotypes e.g. [2]. These observations have usually been incidental by comparing independent studies of different phenotypes and ignoring any correlation between them which was recently thought of as an important factor for power gain. In fact, comparison studies of multivariate and univariate methods usually conclude that multivariate approaches can indeed increase power [3, 4, 5].
\par Regardless of the approach used to analyse multiple phenotypes, population stratification remains an issue when conducting a GWAS. It can generate spurious genotype-phenotype associations i.e. an association that is due to genetic background differences rather than disease status. There are several statistical methods to deal with this issue, among them genomic control and structured association. Genomic control corrects for stratification by modifying the association test statistic at each SNP by a uniform overall factor [6]. However, some SNPs have stronger variation in the allele frequencies across subpopulations than other SNPs. Thus, this uniform adjustment may be insufficient for SNPs with strong differentiation and needless for SNPs with weak differentiation, leading to a loss in power. In structured association, samples are assigned to subpopulation clusters then association testing within each cluster is performed [7]. Among the limitations of this method is the sensitivity to the number of clusters which is not well defined. On the other hand, principal component analysis is considered a fast and an effective way to diagnose population stratification [8]. However, it is not as effective when the samples are related. 
\par Recently, linear mixed-models (LMM) that involve estimating a relatedness matrix have been used and shown to be effective not only in accounting for population stratification but also sample relatedness [9-11]. This is because; given a large number of SNPs it is feasible to make a statement about the relatedness of individuals in a study [12]. Therefore, a well estimated relatedness matrix should provide a complete solution to the problem of population stratification or relatedness. 
\par In addition to GWAS's, heritability; that is the proportion of the phenotypic variance that is due to genetic, is another key application of LMMs in genetics, e.g. [13-15]. Since the emerge of GWASs, estimation of the narrow sense heritability (the variance due to the additive effect) explained by common SNPs via LMM approaches became widely used over classical heritability estimation methods such as regressing offspring phenotype values on the mean parental values. This was motivated by the fact that ideally, heritability is estimated from causal variants [16], which in the context of mixed models means a relatedness matrix from causal SNPs only ($K_{causal}$), However the full set of casual SNPs and their effect sizes are not completely known. Accordingly,  the full set of SNPs in the genotyping platform is used as a proxy for $K_{causal}$ [17].
\par We acknowledge the growing appreciation of the power gained to detect associated SNPs using multivariate approaches over standard univariate analysis, however it is not clear what effect does that has on heritability estimates. Specifically, whether heritability of each phenotype increases as a result of the joint modelling. On the other hand, we also discuss the prospects of using our model for prediction by performing cross validation. To measure the prediction accuracy we use both the root mean square error of prediction and the sample correlation between the predicted and original values.
\par In this study we use the multivariate linear mixed model based on the matrix normal distribution; that  separates the correlations between and within individuals into two components; genetic and environmental. This model was exploited by Zhou and Stephens in their software; GEMMA. While GEMMA is entirely based on the maximum likelihood method, here we adopt Bayesian approaches to estimate the model's parameter with prospects for using this approach for high-dimensional phenotypes; where classical approaches such as the maximum likelihood method fail. We investigate the practical relevance of using this approach in the context of heritability and prediction. Furthermore, we shed a light on the interrelationship between our model and ridge regression. In other words, we provide a general Bayesian interpretation of ridge regression based on the matrix-variate mixed model.
\par We use the Bayesian software JAGS [18] to fit our model through an interface with R called rjags [19]. However, to our knowledge JAGS does not fit the matrix normal distribution directly therefore, the multivariate normal equivalence is used. We then follow Lippert and others [20] and decompose the relatedness matrix which results in independent but not identical standard multivariate normal distributions on the transformed data for each individual. The resulting model contains some scaled covariance matrices which JAGS does not handle, in which case a further simplification is provided. More details are given in the simplified model section.
\par Our results based on heritability estimation and prediction shows that (1) Bayesian estimates form an effective replacement of the standard maximum likelihood estimates. (2) The joint modelling of phenotypes produces more efficient estimates of heritability compared to the univariate analysis. In fact, explained heritability increases significantly under the multivariate analysis. On the other hand,  although our model performs less well for phenotype prediction, we found that imputing the phenotypes first relatively increases the prediction accuracy compared to simply dropping individuals with missing values. All the analysis was performed on a mouse GWAS on two phenotypes from the heterogeneous stock mice data [21].

\section{Methods}
In this section we layout out some definitions and notations about the matrix normal distribution. Next, we discuss the most commonly used mixed model for multiple phenotypes. Then we present the matrix-variate mixed model and its simplified form for JAGS implementation. Finally, we describe the prior distributions used throughout this study.
\subsection{Definitions and notations}
The matrix normal distribution is a generalization of the multivariate normal distribution which allows us to separately model correlations among and within subjects [22]. The probability density function for the random matrix X $(d \times n)$ that follows the matrix normal distribution with mean matrix M $(d\times n)$, column covariance matrix A $(n \times n)$ and row covariance matrix B $(d \times d)$; denoted as $X \sim MN_{n,d} (M, A, B)$ has the form:
\begin{equation}
p(X|M, A, B)=\frac{exp\{\frac{-1}{2}tr[A^{-1}(X-M)B^{-1}(X-M)]\}}{(2\pi)^{nd/2}|A|^{d/2}|B|^{n/2} }
\end{equation}
Its expected value and second order expectations are given by: E[X]=M, $\text{E}[(X-M)(X-M)^t]=B \;\text{tr}(A)$ and $\text{E}[(X-M)^t(X-M)]=A \;\text{tr}(B)$, respectively. \\ 
\\
One way to understand how the matrix normal generalises the multivariate normal distribution is to assume we have n 1-dimensional variates that are independent and identically distributed as normal with zero mean and variance $\sigma^2$ i.e $x_i\sim N(0, \sigma^2)$. This can be written equivalently as a multivariate normal distribution $X_{n\times 1}\sim N_n(0, \sigma^2 I_n)$. Now, assume we have n d-dimensional variate that are independent and identically distributed as multivariate normal with zero mean and covariance matrix B i.e ${x_i}_{d\times 1}\sim(0, B)$. Because the d-dimensional variates are independent, concatenating them will result in a vector with block diagonal covariance matrix $[x_1^t,. . .,x_n^t]\sim N_{nd}(0, I_n\otimes B)$ which is itself equivalent to $[x_1,. . .,x_n]\sim  MN_{n,d} (0, I_n, B)$. Here $C_{n\times n}\otimes D_{d\times d}$ is the Kronecker product defined by

\begin{displaymath}
C\otimes D=\left(
\begin{array}{cccccc}
c_{11}D & c_{12}D &...& c_{1n}D \\
c_{21}D & c_{22}D &...& c_{2n}D \\
. &. &...&.\\
c_{n1}D & c_{n2}D &...& c_{nn}D \\
\end{array}
\right)\\
\end{displaymath}

\subsection{The multivariate linear mixed model}
Assume we have a biallelic SNP and d quantitative phenotypes. The multivariate linear mixed model that involves fixed effects for the SNP effect and random effects to account for correlation within the jth individual is given by:
\begin{equation}
y_{jk}=\beta_k x_j+\eta_{jk}+\epsilon_{jk}\; ,\;\;j=1,2,....,n \;\;, k=1,2,....d
\label{mvLMM}
\end{equation}
\begin{equation}
(\eta_{j1}, \eta_{j2},...,\eta_{jd})\sim N_d(0,\Sigma)\;\;\forall j=1,...n\;\; and \;\; (\epsilon_{j1}, \epsilon_{j2},...,\epsilon_{jd})\sim N_d(0, \Sigma_\epsilon)
\end{equation}
Where n is the number of individuals, d is the number of phenotypes, $y_{jk}$ is the $k^{th}$ component of the d-dimensional phenotypes of the $j^th$ individual. $X_j$ is the genotype of the $j^{th}$ individual at a particular SNP and $\beta_k$ is its effect size for the $k^{th}$ phenotype, $\eta_{jk}$ random effects that are correlated within an individual and independent across different individuals [9].
\subsection{The matrix-variate mixed model}
Model \ref{mvLMM} does not correct for population stratification, for that we exploit the matrix normal distribution that takes into account correlations between individuals in addition to correlation between phenotypes. 
\begin{equation}
Y=\beta X+\eta+\epsilon\;,\;\;\eta\sim \text{MN}_{n,d}(0,K,\Sigma)\;\;and\;\; \epsilon\sim \text{MN}_{n,d}(0,I_{n}, \Sigma_\epsilon)
\label{multi phens that takes into account thier correlation}
\end{equation}
Here, Y is a $d\times n$ phenotypic matrix, X is a $k\times n$ matrix of covariates such as age and sex, $\beta$ is a $d \times k$ matrix of the corresponding coefficients. $\eta$ is a $d\times n$ matrix of random effects that is independent of the $d\times n$ matrix of errors $\epsilon$. The random effect term is used to model any correlation between and within individuals. The $n\times n$ relatedness matrix K represents the genetic covariances between individuals and is typically estimated in advance using the genotype data of p SNPs and n individuals. In other words, it is the sample covariance matrix based on the genotype matrix Z ($p\times n$), with rows pre-processed to have zero mean and unit variance, $K= Z^t.Z / p$. The $d\times d$ matrix $\Sigma$ represents the genetic covariance matrix within individuals. $\Sigma_\epsilon$ and $I_n$ specify the environmental covariance matrices within and between individuals respectively. 
\par This approach, which we use here, does not attempt to test the significance of individual SNPs which GEMMA does by taking a SNP as a fixed effect, instead it provides Bayesian estimates of the narrow sense heritability using an additive model where the phenotype of each individual is defined by a sum of linear effects.

\subsection{Simplified model}
Using univariate LMM, Lippert and others [20] showed that a spectrally transformed model using a spectral decomposition of the relatedness matrix significantly reduces the computational complexity. Similar approaches were adopted later by [5, 10, and 23]. Following these development, we spectrally decompose the relatedness matrix which allows us to write the  matrix-variate mixed model in \ref{multi phens that takes into account thier correlation} as a multivariate LMM on the transformed data for each individual independently as follows:
\begin{equation}
[YU]_{:j}=\beta[XU]_{:j}+\eta_j+\epsilon_j, \eta_j\sim N_d(0,r_j\Sigma)  \;and\; \epsilon_j\sim N_d(0,\Sigma_\epsilon)
\end{equation}
where U is an $n\times n$ orthogonal matrix of normalised eigenvectors and $R=diag(r_1, · · · , r_n)$ is an n by n diagonal matrix filled with the corresponding eigenvalues. Here $[A]_{:j}$ is the $j^{th}$ column of the matrix A. Further, the software JAGS doesn't deal with scaled covariance matrices therefore, we rewrite the model as follows:
\begin{equation}
[YU]_{:j}=\beta[XU]_{:j}+\sqrt{r_j}\zeta_j+\epsilon_j,  \;\zeta_j\sim N_d(0,\Sigma) \;and \;\epsilon_j\sim N_d(0,\Sigma_\epsilon) 
\end{equation}

\label{simplified Matrix-variate mixed model}
\subsection{Priors}
JAGS is a fully Bayesian software, meaning that we need to assign a prior distribution to each parameter. In addition, the multivariate normal distribution in the BUGS language is parameterised in terms of its mean and precision (the inverse of the variance-covariance matrix). Accordingly, we assign the conjugate prior of the precision matrix; that is the Wishart distribution to both $\Sigma^{-1}$ and ${\Sigma_\epsilon}^{-1}$ [24]. For the covariates' coefficients we place a diffuse prior in the form of a normal distribution such as a multivariate normal distribution with mean zero and large covariance matrix e.g. $\tilde{\Sigma}=diag(.0001, d)$.
\par The Wishart prior $W_d(V, \nu)$ is characterised by a scalar degrees of freedom $\nu>d-1$ and a location(scaling) matrix V; that has the same dimension as the underlying covariance matrix ($d\times d$). The location matrix, as it is named possesses information about the location of each element of the underlying covariance matrix whereas the degrees of freedom reflect the strength of beliefs in the location values [25].  Therefore, to assign a diffuse (non-informative) Wishart prior, a few degrees of freedom has to be set such that the Wishart distribution ramains proper. Thus, setting $\nu=d$ is a common choice. On the other hand, V is usually chosen as the maximum likelihood estimate or the identity matrix depending on the amount of shrinkage one wants to impose as well as the size of the sample [26].
\par In small GWASs with too many parameters to be estimated, the prior can be quite influential.  However, although the data we are analysing in this paper is considered to be a small GWAS data (n=1940 and $p\sim 12000$) the number of parameters to be estimated is relatively small as only two phenotypes have been selected. Accordingly, the choice of the identity matrix as a scaling matrix to the diffuse Wishart prior is considered very effective [26]. Nevertheless, to see the effect of different prior specification we use the maximum likelihood estimates from GEMMA as a scaling matrix in addition to the identity matrix.
\par
\begin{figure}[H]
\begin{center}
\frame{\includegraphics[width=3 in]{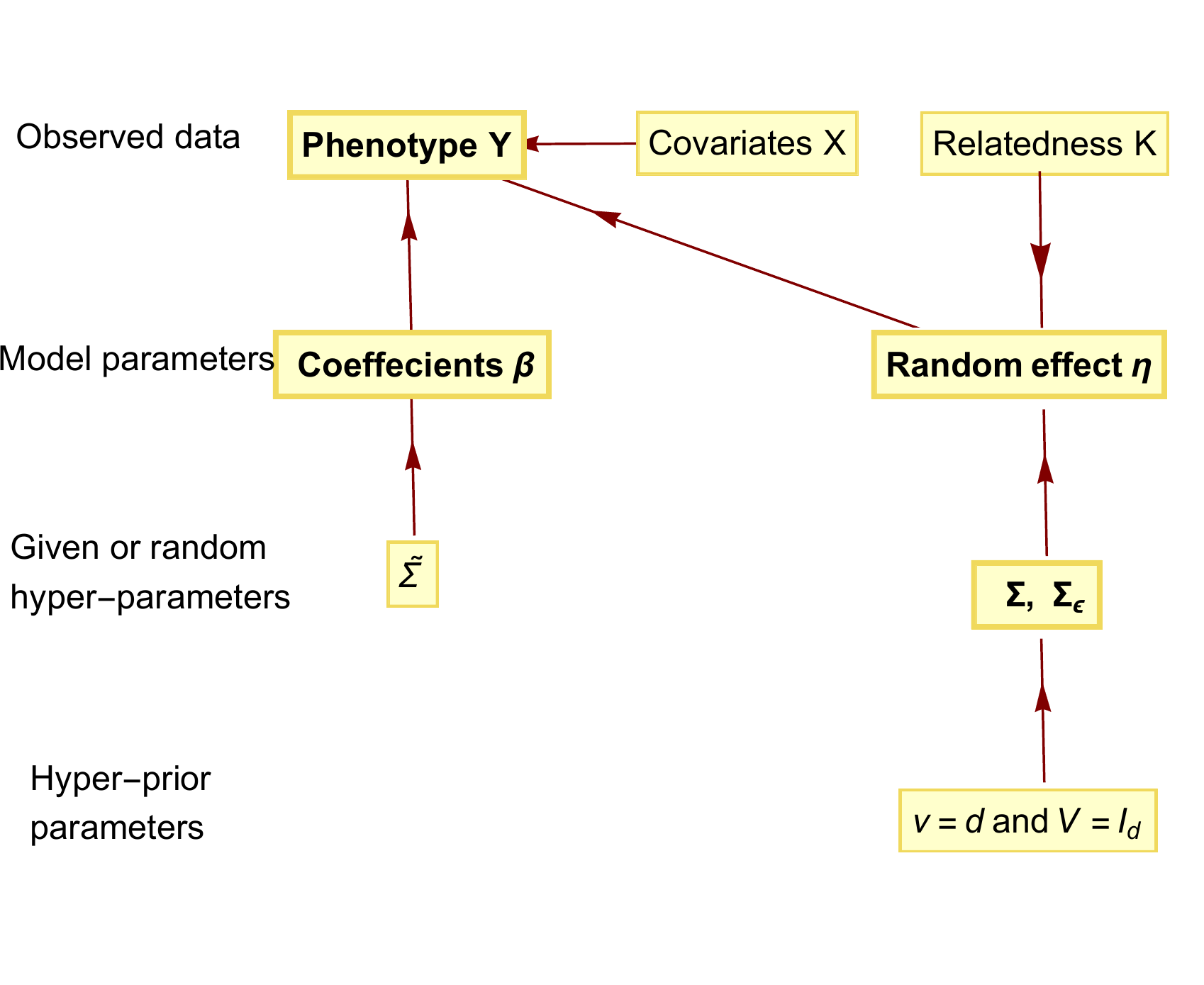}}
\end{center}
\caption{Hierarchical structure of the model. The bold represent random variables otherwise fixed parameters }
\label{tree}
\end{figure}

 To illustrate our approach, we present the hierarchical model depicted in figure \ref{tree}. It has a total of four layers; the observed data is located in the first layer and contains phenotype and genotype information, plus a known relatedness matrix. The second layer contains the covariates' coefficients $\beta$ as well as the random effect parameters $\eta$. The third layer comprises the hyper-parameters; $\Sigma$ and $\Sigma_\epsilon$ which are assumed to be random matrices with Wishart hyper-priors whereas $\tilde{\Sigma}$ is assumed to be fixed. Finally, the fourth layer contains the hyper-prior parameters, namely the degrees of freedom and the scaling matrix which are both fixed and equal to the number of phenotypes d and the identity matrix, respectively.

\section{Generalised Bayesian interpretation of ridge regression}
The Bayesian prospective of ridge regression assumes that the regression coefficients of a multivariate  regression are independent and identically (iid) normally distributed [27]. Here we aim to give a broader Bayesian interpretation of ridge regression in the context of matrix normal distribution.
Consider the matrix normal regression model of p SNP effects on d phenotypes:
\begin{equation}
Y=\beta_z Z+\epsilon\;,\;\; \epsilon\sim \text{MN}_{n,d}(0,I_{n}, \Sigma_\epsilon)
\label{simple mv model}
\end{equation}
with matrix normal prior on the effect sizes:
\begin{equation}
\beta_z\sim MN(0, I_p, \Sigma_\beta)
\label{MN prior on beta}
\end{equation}
where the $I_p$ ($p\times p$) and $\Sigma_\beta$ ($d\times d$) represent the effect size covariances between and within SNPs. This means we are assuming that effect sizes are correlated within SNPs and independent across SNPs with unity variance. Exploiting the multivariate normal equivalence of matrix normal distribution, model $\ref{simple mv model}$ can be rewritten as:
\begin{equation}
Vec(Y)=Z^t\otimes I_d Vec(\beta_z)+Vec(\epsilon);\;\;Vec(\epsilon)\sim N_{nd}(I_n\otimes\Sigma_\epsilon)
\label{N prior on vec beta}
\end{equation}
Similarly the prior on the effect sizes:
\begin{equation}
Vec(\beta_z)\sim N_{dp}(0, I_p\otimes \Sigma_\beta)\;,
\end{equation}
which is itself equivalent to $[\beta]_{:j}\sim N_d(0, \Sigma_\beta)$, j=1,...,p.\\ 
Now, 
\begin{dmath}
V(Z^t\otimes I_d Vec(\beta_z))=(Z^t\otimes I_d)(I_p\otimes \Sigma_\beta)(Z^t\otimes I_d)^t\\
=(Z^t \otimes \Sigma_\beta)(Z^t\otimes I_d)^t\\
=(Z^t \otimes \Sigma_\beta)(Z\otimes I_d)\\
=Z^tZ\otimes\Sigma_\beta
\end{dmath}
Recall the multivariate normal equivalence of our model without the fixed effect term:
\begin{equation}
Vec(Y)=Vec(\eta)+Vec(\epsilon); Vec(\eta)\sim N_{nd}(K\otimes\Sigma),  Vec(\epsilon)\sim N_{nd}(I_n\otimes\Sigma_\epsilon), 
\label{no_fixed_effect}
\end{equation}
It is clear that our model is equivalent to the matrix normal regression with matrix normal prior on the effect sizes when the relatedness matrix is estimated using the available SNPs i.e $K=Z^t Z/p$. 
\label{relations}

\section{Application}
Here we analyse a mouse data from the heterogeneous stock mice data [21]. It is a small GWAS data set with 2 phenotypes: the percentage of cluster of differentiation (CD8+) of the cells with no measurements in 27\% of the individuals and the mean corpuscular haemoglobin (MCH) with no measurements in 18\% of the individuals. The phenotypes were already corrected for sex, age, body weight, season and year effects by the original study. Also the data has been quantile normalised to a standard normal distribution. On the other hand, we have a total of 12,226 autosomal SNPs, with missing genotypes replaced by the mean genotype of that SNP. 
\subsection{JAGS implementation}
\par We use the rjags package to fit the model described in section \ref{simplified Matrix-variate mixed model}. It provides an interface from R to the JAGS library and uses Markov Chain Monte Carlo (MCMC) to generate a sequence of dependent samples from the posterior distribution of the parameters to be estimated. 
\par In order to monitor convergence we run several chains for a number of cycles (burn-in) so that the model will reach a stable state. In other words, the chains should converge to the target distribution, namely the joint posterior distribution of the model's parameters. For convergence diagnostics and samples' summary we use the package coda [28] which is designed for analysing MCMC output.
\par
\begin{figure}[H]
\begin{center}
\frame{\includegraphics[width=3.7in]{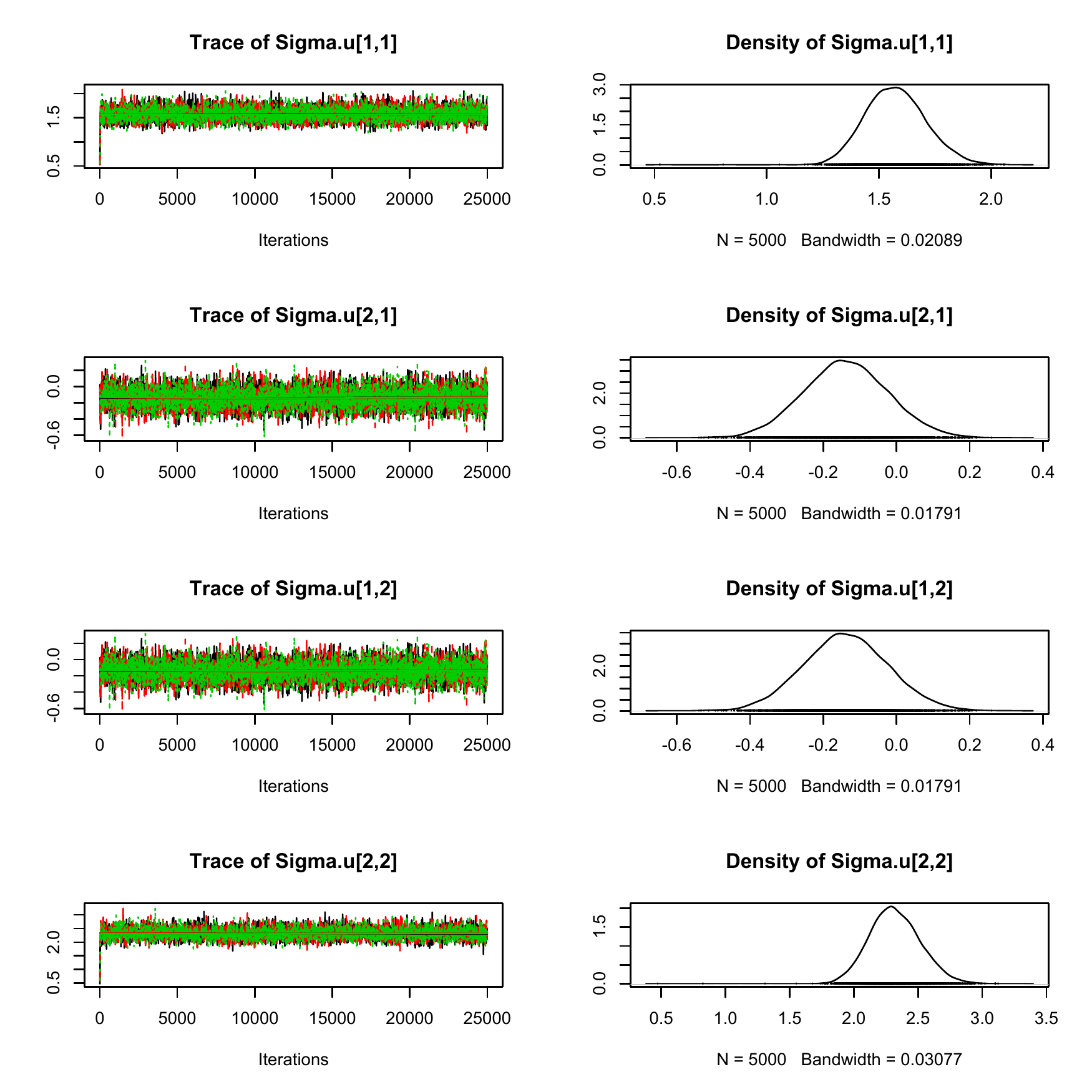}}
\end{center}
\caption{Trace and density plots for $\Sigma$}
\label{density_plots_sigma.u}
\end{figure}

\begin{figure}[H]
\begin{center}
\frame{\includegraphics[width=3.7 in]{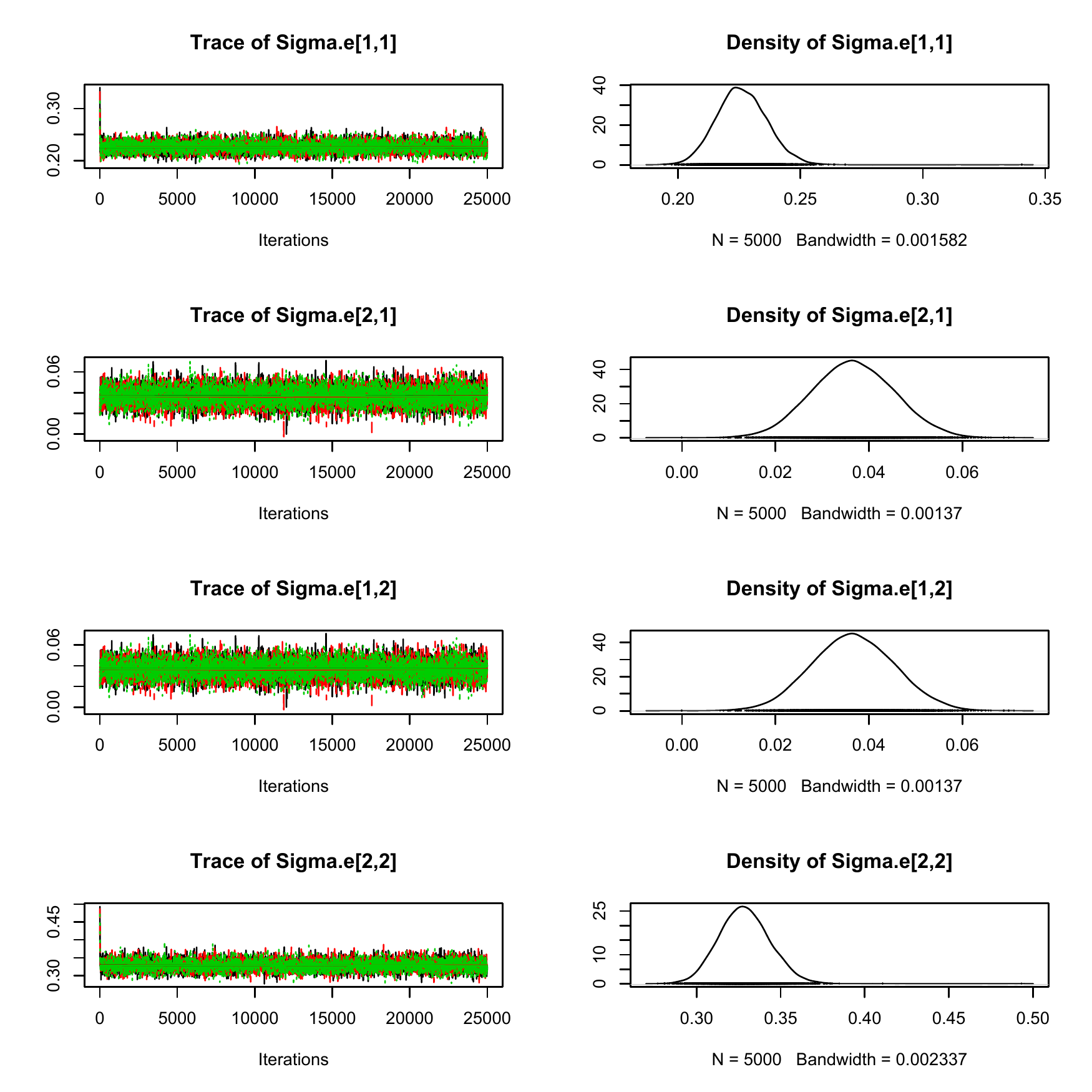}}
\end{center}
\caption{Trace and density plots for $\Sigma_\epsilon$}
\label{density_plots_sigma.e}
\end{figure}

 Using the prior distributions and the input values given in figure \ref{tree}, we run 35000 iterations of three chains, with the first 10,000 discarded. Then we sub-sample every 5th value of the parameter to be estimated, giving from each chain a sample size of 5000 from the posterior distribution. Rjags and coda outputs for the both $\Sigma$ and $\Sigma_\epsilon$ are shown below. Table \ref{bayes_estimates} shows the Bayesian estimates (posterior means), standard deviation, naive standard error which ignores autocorrelation of the chain, times series SE which takes that correlation into account and finally the credible intervals for both $\Sigma$ and $\Sigma_\epsilon$. On the other hand, figures \ref{density_plots_sigma.u}, \ref{density_plots_sigma.e} are given to heuristically shows that the number of iterations used was sufficient to produce acceptable convergence, as all the chains appear to be overlapping one another.

\begin{table}[H]
\caption{Bayesian estimates of $\Sigma$ and $\Sigma_\epsilon$, their standard error and credible intervals}
\centering
\scalebox{0.85}{
\begin{tabular}{c c c c c c c c c c c}
\hline
&Mean& SD & Naive SE& Time-series SE& 2.5\% & 25\%& 50\% &75\%& 97.5\% \\ [0.5ex] 
\hline
\hline
&&&&\\
${\Sigma_\epsilon}_{11}$ & 0.23 & 0.0102 & 0.0006 & 0.0006 & 0.21 & 0.22 & 0.22  & 0.23& 0.25 \\
&&&&\\
${\Sigma_\epsilon}_{12}$ & 0.04 & 0.0088 & 0.0005 & 0.0005 & 0.02 & 0.03 & 0.04 & 0.04 & 0.05 \\
&&&&\\
${\Sigma_\epsilon}_{22}$ & 0.33 & 0.0152 & 0.0009  & 0.0009 & 0.02 & 0.03 &0.04 & 0.04  &0.05 \\
&&&&\\
${\Sigma}_{11}$ & 1.57 & 0.1399 & 0.0081 &  0.0109 & 1.34 & 1.47 & 1.57 & 1.67 & 1.86 \\
&&&&\\
${\Sigma}_{12}$ & -0.13 & 0.1147 & 0.0066 & 0.0009 & -0.34 & -0.21 & -0.13 & -0.06 & 0.11 \\
&&&&\\
${\Sigma}_{22}$ & 2.33 & 0.2009 & 0.0116 & 0.0125 & 1.94 & 2.2 & 2.32 & 2.45 & 2.81 \\
\hline
\end{tabular}}
\label{bayes_estimates}
\end{table}

\subsection{Heritability estimation}
In our model, the narrow sense heritability is defined as the ratio of the diagonal entries of $\Sigma$ to the sum of the diagonal entries of $\Sigma$ and $\Sigma_\epsilon$. Therefore the heritability of phenotype i is: $h_i=\frac{\Sigma_{ii}}{\Sigma_{ii}+{\Sigma_\epsilon}_{ii}}$. 
\par Table \ref{h} shows estimates of the narrow sense heritability as well as their standard error using three different approaches :(1) Bayesian multivariate analysis using Jags, (2) Classical multivariate analysis using GEMMA. (3) Classical univariate analysis using GEMMA. To see the effect of imputation on heritability, each of the above approaches were taken first after dropping individuals with missing phenotypes which results in 1197 individuals being analysed. Second after an imputation step using the best linear unbiased predicator described in the next section. 
\par From table \ref{h}, we can see that there is $\sim 25-30\%$ increase in the explained heritability when phenotypes were jointly modelled. In fact based on their standard error, the joint modelling provides more accurate estimates of the heritability over the univariate modelling. On the other hand, there is $\sim 2.5\%$ increase in  heritability using Bayesian estimates as opposed to the maximum likelihood estimates. Overall, we see that imputing phenotypes first provides higher estimates of heritability. Here the reported standard errors are the ones that take the chain's correlation into account using estimates of the spectral density at zero [28]; which is usually greater than the naive standard error. It is worth noting that after imputation, heritability of both phenotypes were almost the same, however when individuals with missing phenotype were dropped, heritability of CD8+ is greater than heritability of MCH, which could be due to the fact that the percentage of the missing values in CD8+ is higher. 

\begin{table}[H]
\caption{Heritability estimates and their standard error. $h_1$ and $h_2$  are narrow sense heritability estimates of the percentage of CD8+ cells and the mean corpuscular haemoglobin (MCH), respectively. Here, Bayesian mv-JAGS uses Bayesian estimates of the matrix-variate mixed model's parameters obtained from JAGS. mv-GEMMA uses maximum likelihood estimates of the Matrix-variate mixed model’s parameters obtained from GEMMA. Univ-GEMMA uses maximum likelihood estimates based on a univariate LMM.}\centering
\scalebox{0.85}{
\begin{tabular}{@{}rrrcrrcrr@{}}\toprule
& \multicolumn{2}{c}{Baysian mv-JAGS} & \phantom{abc}& \multicolumn{2}{c}{mv-GEMMA}& \phantom{abc}&\multicolumn{2}{c}{univ-GEMMA} \\
 \cmidrule{2-3} \cmidrule{5-6} \cmidrule{8-9} 
&$h_1 (SE)$ & $h_2$(SE) &&$h_1$ & $h_2$ && $h_1$ (SE)& $h_2$ (SE) \\ \midrule
Without imputation & 0.82 (0.0003) & 0.85(0.0003)&& 0.8  & 0.83 && 0.61 (0.03) &  0.64 (0.03) \\
With imputation & 0.87 (0.0009)& 0.88 (0.0009)&& 0.86 & 0.86&&0.69(0.02) & 0.7 (0.02) \\

 \bottomrule
\end{tabular}}
\label{h}
\end{table}

\subsection{Prediction as model checking}
\subsubsection*{Best linear unbiased predictor}
Here we use cross validation to see how well our model predicts the phenotype. For this we partition the data into complementary subsets ($Y_1,Y_2$); training and validating sets. We assess any conflict between the observed data in the validating set and their predictive values from the training set using the root mean square error matrix $RMSE=[(Y_2-\hat{Y_2})(Y_2-\hat{Y_2})^t]/n_2$ as well as the sample correlation. \\
\\
Using the equivalence between the matrix normal and multivariate normal, equation \ref{multi phens that takes into account thier correlation} can be rewritten as:
\begin{equation}
Vec(Y)=X^t\otimes I_d Vec(\beta)+Vec(\eta)+Vec(\epsilon); Vec(\eta)\sim N_{nd}(0, K\otimes\Sigma),  Vec(\epsilon)\sim N_{nd}(0, I_n\otimes\Sigma_\epsilon), 
\end{equation}
which imply $Vec(Y)\sim N_{nd} (\mu, H)$
Where $\mu=X^t\otimes I_d Vec(\beta)$ and $H=K\otimes\Sigma+I_n\otimes\Sigma_\epsilon$. Next, we partition the mean vector and covariance matrix to perform cross validation as follows:
$\mu=[\mu_o, \mu_m]^t $and $H=\left(
\begin{array}{cc}
H_{oo} & H_{om} \\
H_{om} & H_{mm} \\
\end{array}
\right)$,
where the subscripts o and m refer to observed and missing respectively. Then, it follows that $Vec(Y_m)|Vec(Y_o)$ is normally distributed with mean:
\begin{equation}
\widehat{Vec(Y_m)}=\mu_m+H_{mo} H_{oo}^{-1}(Vec(Y_o)-\mu_o)
\end{equation} 
\par As in heritability estimation, we want to see what effect imputation has on prediction accuracy. Accordingly, the cross validation is performed (1) after an imputation step for the missing phenotypes using GEMMA  (2) after simply dropping individual with missing phenotypes. In each scenario three approaches are used to predict; (1) GEMMA which will deal with this as an imputation step using BLUP with maximum likelihood estimates. (2) BLUP with Bayesian estimates using the identity matrix as a hyper-prior parameter for $\Sigma$ and $\Sigma _\epsilon$ (3) Also BLUP but with a hyper-prior parameter equal to the inverse of the mle's from GEMMA. The last approach is taken to see the effect of prior specification on prediction accuracy.
\par Table \ref{pred} shows that the performances of BLUP based on maximum likelihood estimates and bayesian estimates using different prior specification are very similar with average correlations (RMSE) 0.56 (0.79) and 0.67 (0.69) before and after imputation, respectively. This means that the imputation step increased the average prediction accuracy. 
\begin{table}[H]
\caption{Average RMSE and correlation}\centering
\scalebox{0.8}{
\begin{tabular}{@{}rrrrrcrrr@{}}\toprule
\multicolumn{3}{c}{Dropping } & \phantom{abc}& \multicolumn{3}{c}{ Imputation} \\
 \cmidrule{2-4} \cmidrule{6-8} 
&GEMMA & Scaling=Identity & Scaling=mle && GEMMA & Scaling=Identity & Scaling=mle\\
RMSE & 0.78 &  0.79 & 0.79 && 0.69 & 0.69 &  0.69   \\
Corr & 0.58 & 0.58 & 0.53 && 0.67 & 0.67 & 0.67 \\
 \bottomrule
\end{tabular}}
\label{pred}
\end{table}

\subsection{Effect size distribution}
\label{dist}
\begin{figure}[H]
\begin{center}
\mbox{\subfigure[Smoothed histogram of the effect sizes reported using GEMMA.]{\frame{\includegraphics[width=2.75in]{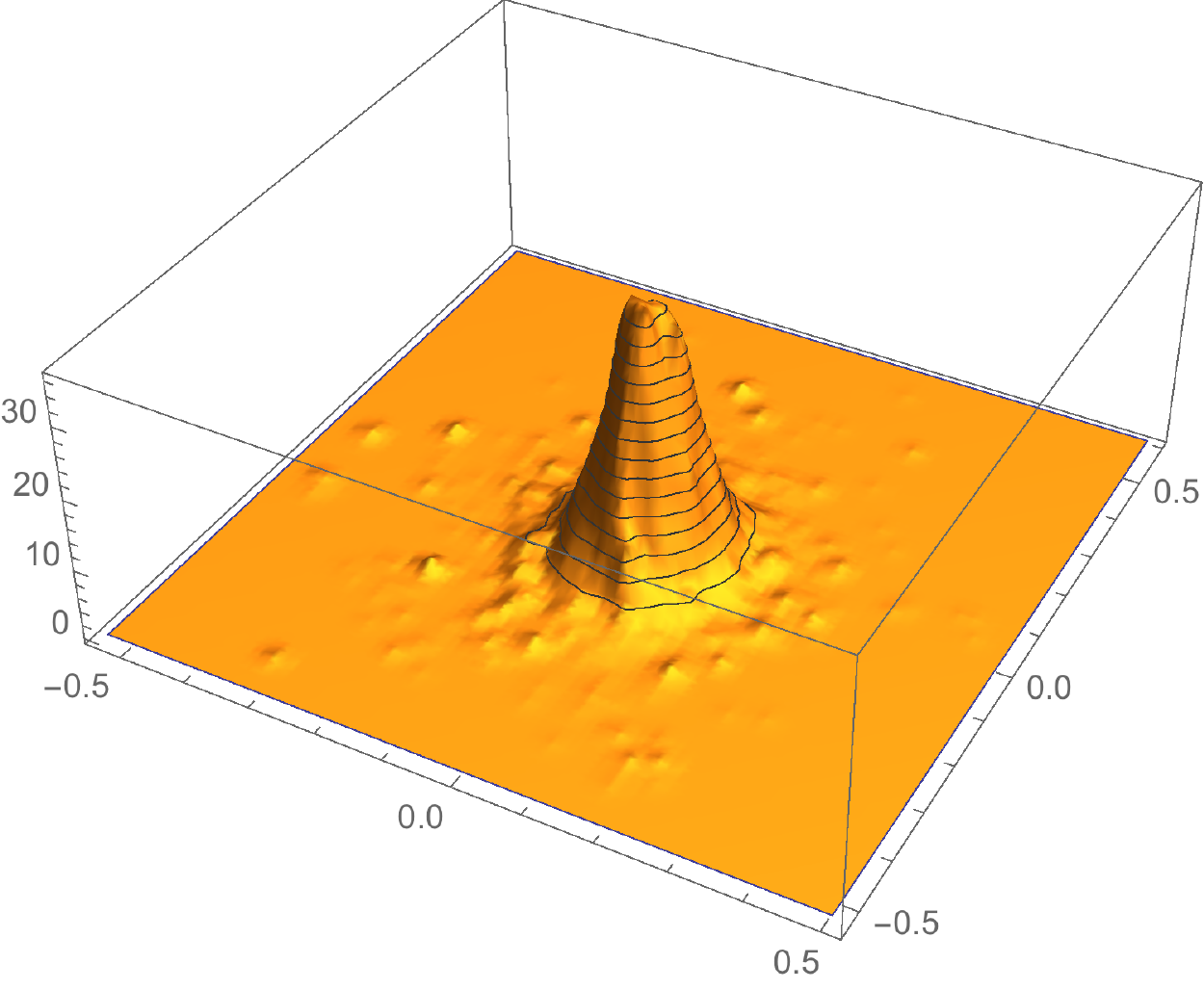}}}\quad
\subfigure[prior distribution of the ffect sizes based on our model. For this we generated a matrix from the Wishart distribution with identity scaling matrix and two degrees of freedom.]{\frame{\includegraphics[width=2.9in]{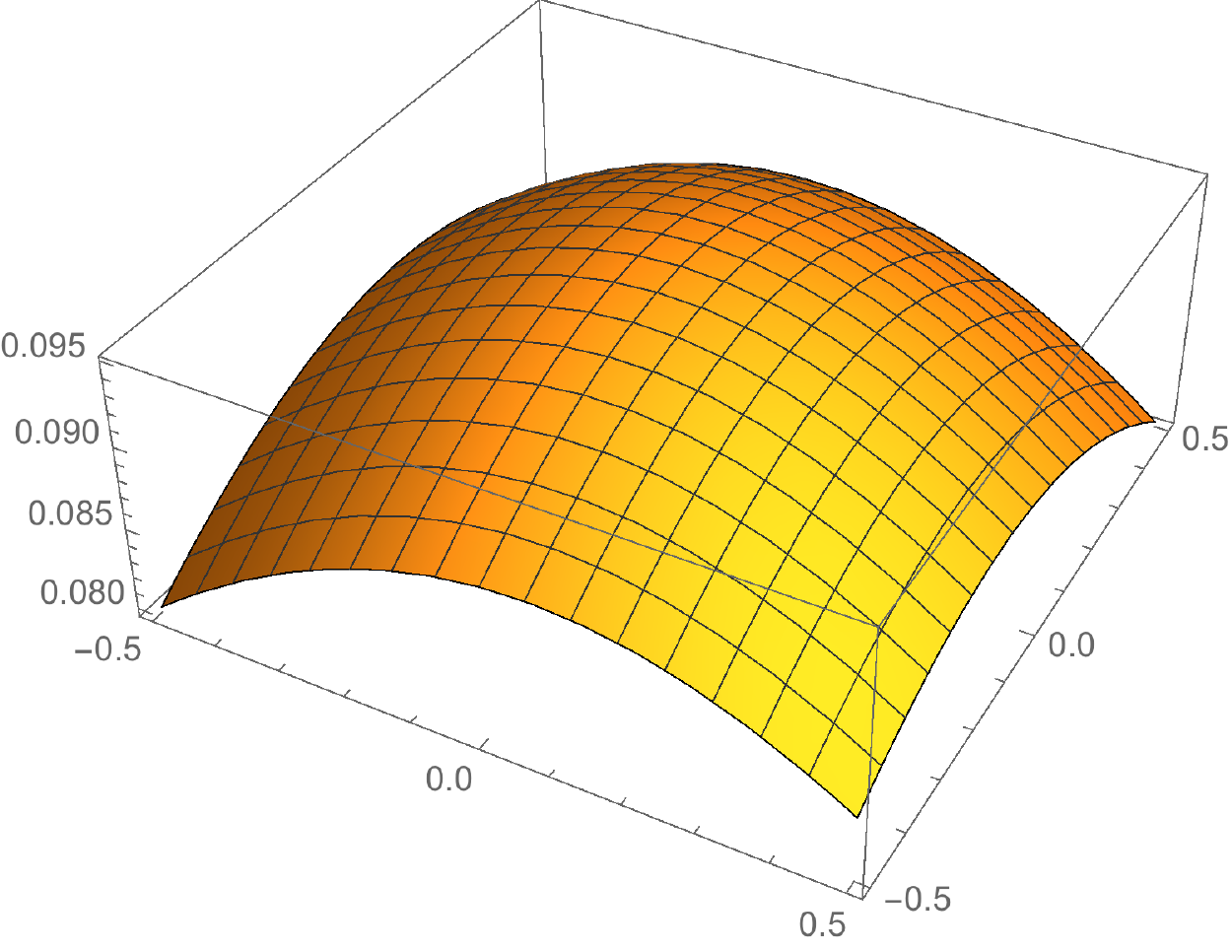}}}}
\end{center}
\caption{Effect size distributions}
\label{smoothed_histograms}
\end{figure}

It is known that phenotype prediction depends on the distribution of the effect sizes. Here we examine the prior distribution of the effect sizes in an attempt to explain the lack of prediction accuracy (based on RMSE). Figure \ref{smoothed_histograms} a and b show the distribution of the effect sizes given by GEMMA and the prior distributions we used; that is $[\beta]_{:j}\sim N_d(0, \Sigma_\beta)$, j=1,...,p with $\Sigma_\beta$ distributed as inverse wishart with identity scaling matrix and two degrees of freedom. It is clear from the figures that this distribution significantly overestimates the number of SNPs with large effect sizes which does not reflect our prior understanding that most SNPs have negligible effect. We believe this is a potential explanation why prediction is compromised in our model.
\par
\begin{figure}[H]
\begin{center}
\frame{\includegraphics[width=3.in]{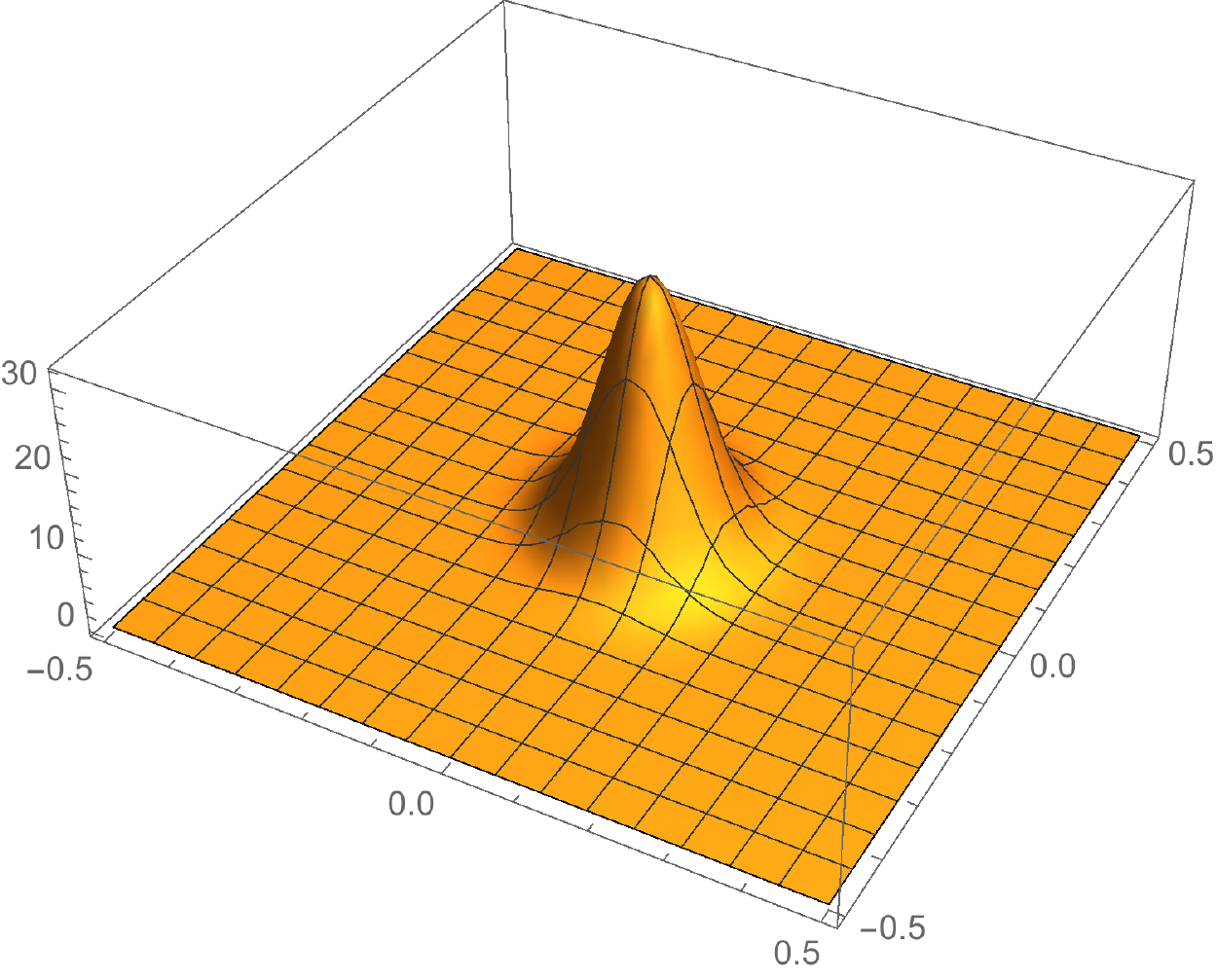}}
\end{center}
\caption{Density function of the effect sizes with rescaled relatedness matrix (0.003K)}
\label{Density_function}
\end{figure}

 Recall that our model assumes that effect sizes are independent across SNPs and all have unity variance (see section \ref{relations}) which is not necessarily the case. If we modify the model replacing $I_p$ by $\sigma^2_\beta I_p$; in other words we are assuming that SNPs are homogeneous in the sense that their effect sizes have the same variance $\sigma^2_\beta \neq 1$, then 
\begin{dmath}
V(Z^t\otimes I_d Vec(\beta))=\sigma^2_\beta Z^tZ \otimes\Sigma_\beta
\end{dmath}
which is equivalent to equation \ref{no_fixed_effect} with a rescaled relatedness matrix $K\rightarrow\sigma^2_\beta K$. It appears that a much smaller value of $\sigma^2_\beta$ such as 0.003  reflects the prior knowledge that most of the SNPs are null better than $\sigma^2_\beta=1$, see figure \ref{Density_function}. 

\section{Discussion}
In this study we showed that the joint modelling of the phenotypes; CD8+ and MCH using a matrix-variate mixed model that takes into account both genetic correlation between phenotypes as well as between individuals, provides 25-30\% increase in the explained heritability. In fact, based on the standard error, heritability estimates using our model were more efficient as opposed to the univariate modelling where heritability of each phenotype is estimated separately. In addition, we showed that in the mouse data which has a high percentage of missing values (27\% and 18\% for CD8 and MCH), an initial imputation step increased the average prediction accuracy.
\par At the beginning of this study it was not clear to us wether some SNPs can be taken as covariates (fixed effect) in addition to the relatedness matrix until we showed that when the relatedness matrix is estimated using $K=Z^tZ/p$ from the genome- wide SNPs, the matrix-variate mixed model is in fact a multi-snp model with a matrix normal prior on the effect sizes. Because this matrix normal appears to have an identity matrix for the between effect sizes covariances, we viewed this as a generalisation of the known Bayesian interpretation of ridge regression where effect sizes are assumed to be iid normal.
\par We provided a simplified form of the matrix-variate mixed model which allows fitting it using many "off-the-shelf" Bayesian software. One of the conclusions drawn is that the rjags package performs excellently among various competitors. This is very encouraging as it saves the user having to write their own MCMC code. However, it should be noted that although JAGS perform perfectly for heritability estimation and prediction, alternative software might be needed for a genome- wide association scan, as JAGS will be intrinsically slow, due to the number of iterations required by MCMC in order for the chain to converge.
\par In both heritability estimation and prediction, we observed similar results using either maximum likelihood estimates from GEMMA or Bayesian estimate from our hierarchical model. This means that our Bayesian estimates can seamlessly be used in place of the traditional maximum likelihood estimates from GEMMA. The usefulness of this conclusion accentuates when heritability estimation is the interest using high-dimensional phenotypes where the maximum likelihood method fails due to lack of information (small sample size) to efficiently estimate the model’s parameters or due to the positive definiteness constraint on the covariance matrices which is numerically problematic regardless of the available amount of information. Our Wishart hyper-prior will eliminate these concerns.
\par Although high-dimensional phenotypes is an appealing domain of application for the hierarchical model we proposed, the choice of the Wishart scaling matrix can be quite critical. This is because of the restriction on the Wishart distribution to be proper; that is the df be $>d-1$. Clearly, the severity of this restriction increases significantly with d (number of phenotypes), as it forces the Wishart to remain somewhat informative rather than very diffuse. Currently, we are exploring different prior specifications using gene expression data from the TwinsUK cohort. Specifically, Hierarchical Wishart that has an unknown diagonal scaling matrix of the form $V=aI$ with unknown degrees of freedom $\nu$. To estimate a and $\nu$ we choose to add an extra level of variability by placing a flat prior on a and similarly on $\nu$, remembering to set its value to be always greater than d-1, so that the Wishart distribution remains proper. The use of this prior form is equivalent to the use of the rescaled relatedness matrix described previously.
\par Regarding prediction, as we mentioned in section \ref{dist}, our model assumes equal effect size variances which can be inadequate for large, heterogeneous regions. Expanding the model to account for different classes of SNPs with distinct effect size variances can improve prediction [29].

\begin{section}*{References}
\addcontentsline{toc}{section}{References}
\begin{easylist}
& Teslovich, Tanya M., et al. "Biological, clinical and population relevance of 95 loci for blood lipids." Nature 466.7307 (2010): 707-713.\\
& Sivakumaran, Shanya, et al. "Abundant pleiotropy in human complex diseases and traits." The American Journal of Human Genetics 89.5 (2011): 607-618.\\
& Stephens, Matthew. "A unified framework for association analysis with multiple related phenotypes." PloS one 8.7 (2013): e65245.\\
& O’Reilly, Paul F., et al. "MultiPhen: joint model of multiple phenotypes can increase discovery in GWAS." PLoS One 7.5 (2012): e34861.\\
& Zhou, Xiang, and Matthew Stephens. "Efficient multivariate linear mixed model algorithms for genome-wide association studies." Nature methods 11.4 (2014): 407-409.\\
& Devlin, B., and Kathryn Roeder. "Genomic control for association studies." Biometrics 55.4 (1999): 997-1004.\\
& Pritchard, Jonathan K., et al. "Association mapping in structured populations." The American Journal of Human Genetics 67.1 (2000): 170-181.\\
& Price, Alkes L., et al. "Principal components analysis corrects for stratification in genome-wide association studies." Nature genetics 38.8 (2006): 904-909.\\
& Yang, Qiong, and Yuanjia Wang. "Methods for analyzing multivariate phenotypes in genetic association studies." Journal of probability and statistics 2012 (2012).\\
& Zhou, Xiang, and Matthew Stephens. "Genome-wide efficient mixed-model analysis for association studies." Nature genetics 44.7 (2012): 821-824.\\
& Korte, Arthur, et al. "A mixed-model approach for genome-wide association studies of correlated traits in structured populations." Nature genetics 44.9 (2012): 1066-1071.\\
& Balding, David J. "A tutorial on statistical methods for population association studies." Nature Reviews Genetics 7.10 (2006): 781-791.\\
& Yang, Jian, et al. "Common SNPs explain a large proportion of the heritability for human height." Nature genetics 42.7 (2010): 565-569.\\
& Grundberg, Elin, et al. "Mapping cis-and trans-regulatory effects across multiple tissues in twins." Nature genetics 44.10 (2012): 1084-1089.\\
& Zaitlen, Noah, and Peter Kraft. "Heritability in the genome-wide association era." Human Genetics 131.10 (2012): 1655-1664.\\
& Donnelly, Peter. "Progress and challenges in genome-wide association studies in humans." Nature 456.7223 (2008): 728-731.\\
& Hayes, Ben John, Peter M. Visscher, and Michael E. Goddard. "Increased accuracy of artificial selection by using the realized relationship matrix." Genetics Research 91.01 (2009): 47-60.\\
& Plummer, M. (2003), “JAGS: A Program for Analysis of Bayesian Graph- ical Models Using Gibbs Sampling,” in Proceedings of the 3rd International Workshop on Distributed Statistical Computing, Vienna, p. 125. [128].\\
& Plummer, Martyn, et al. "Package ‘rjags’." update 16 (2015): 1.\\
& Lippert, Christoph, et al. "FaST linear mixed models for genome-wide association studies." Nature Methods 8.10 (2011): 833-835.\\
& Valdar, William, et al. "Genome-wide genetic association of complex traits in heterogeneous stock mice." Nature genetics 38.8 (2006): 879-887.\\
& Dawid, A. Philip. "Some matrix-variate distribution theory: notational considerations and a Bayesian application." Biometrika 68.1 (1981): 265-274.\\
& Pirinen, Matti, Peter Donnelly, and Chris CA Spencer. "Efficient computation with a linear mixed model on large-scale data sets with applications to genetic studies." The Annals of Applied Statistics 7.1 (2013): 369-390.\\
& Gelman, Andrew, et al. Bayesian data analysis. Vol. 2. London: Chapman $\&$ Hall/CRC, 2014.\\
& Sinay, Marick S., Chi-Wen Hsu, and John SJ Hsu. "Bayesian estimation with flexible prior for the covariance structure of linear mixed effects models." International Journal of Statistics and Probability 2.4 (2013): p29.\\
& Daniels, Michael J., and Robert E. Kass. "Nonconjugate Bayesian estimation of covariance matrices and its use in hierarchical models." Journal of the American Statistical Association 94.448 (1999): 1254-1263.\\
& Hoerl, Arthur E., and Robert W. Kennard. "Ridge regression: Biased estimation for nonorthogonal problems." Technometrics 12.1 (1970): 55-67.\\
& Plummer, Martyn, et al. "Package ‘coda’." (2015).\\
& Speed, Doug, and David J. Balding. "MultiBLUP: improved SNP-based prediction for complex traits." Genome research 24.9 (2014): 1550-1557.
\end{easylist}
\end{section}
\section{Appendix}
\begin{figure}[H]
\hspace*{-2cm}  
\includegraphics[width=7 in]{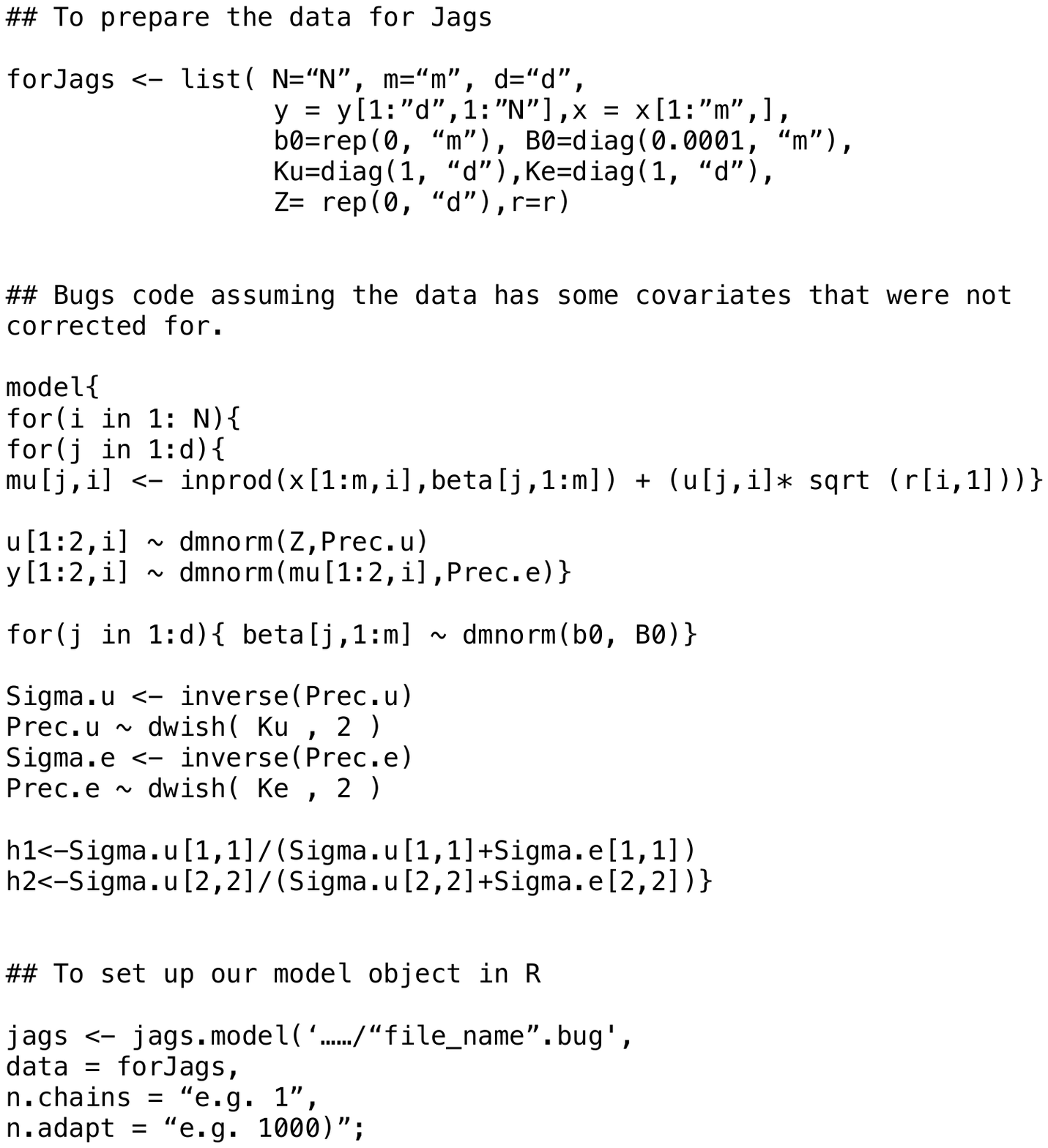}
\end{figure}

\end{document}